\documentclass[preprint2]{aastex}
\received{Aug 4 2004}
\date{Aug 4 2004}
\accepted{}
\slugcomment{}
\shorttitle{Fundamental Plane Evolution of AGN Hosts}
\shortauthors{Woo et al.}

\begin{document}

\title{The Fundamental Plane Evolution of Active Galactic Nucleus Host Galaxies}

\author{Jong-Hak Woo\altaffilmark{1,2,3},
C. Megan Urry\altaffilmark{4},
Paulina Lira\altaffilmark{5},
Roeland P. van der Marel\altaffilmark{6},
Jose Maza\altaffilmark{7}}

\altaffiltext{1}{Department of Astronomy, Yale University, P.O. Box 208101, New Haven, CT 06520-8101; jhwoo@astro.yale.edu}
\altaffiltext{2}{Visiting Astronomer, Kitt Peak National Observatory, National Optical Astronomy Observatory, which is operated by the Association of Universities for Research in Astronomy, Inc. (AURA) under cooperative agreement with the National Science Foundation} 
\altaffiltext{3}{Visiting Astronomer, Cerro Tololo Inter-American Observatory, which is operated by the AURA, Inc., under agreement with the National Science Foundation} 
\altaffiltext{4}{Department of Physics and Yale Center for Astronomy and Astrophysics, Yale University, P.O. Box 208121, New Haven, CT 06520-8121; meg.urry@yale.edu}
\altaffiltext{5}{Departamento de Astronomía, Universidad de Chile, Casilla 36-D, Santiago, Chile; plira@das.uchile.cl}
\altaffiltext{6}{Space Telescope Science Institute, 3700 San Martin Dr. Baltimore MD 21218; marel@stsci.edu                      }
\altaffiltext{7}{Departamento de Astronomía, Universidad de Chile, Casilla 36-D, Santiago, Chile; jmaza@das.uchile.cl}

\begin{abstract}
We measured the stellar velocity dispersions of 15 active galactic nucleus (AGN) host
galaxies at redshifts as high as $\sim 0.34$. 
Combining these with published velocity dispersion measurements from the literature,
we study the Fundamental Plane of AGN host galaxies and its evolution.
BL Lac hosts and radio galaxies seem to lie on the same Fundamental
Plane as normal early-type galaxies.
The evolution of the mass-to-light ratio of AGN host galaxies shows a similar trend to
that observed in normal early-type galaxies, 
consistent with single-burst passive evolution models with formation redshifts $z \gtrsim 1$.
The lack of a significant difference between normal and AGN host galaxies in the Fundamental plane
supports the ``Grand Unification" picture wherein AGNs are a transient phase in the evolution of normal galaxies.
The black hole masses of BL Lac objects and radio galaxies, derived using the mass -- dispersion relation,
are similar. The black hole mass is independent of BL Lac type.
The local black hole mass -- host galaxy luminosity relation of our sub-sample at $z < 0.1$ is similar to
that of local normal and radio galaxies, but is less well defined at higher redshift
due to the luminosity evolution of the host galaxies.
\end{abstract}

\keywords{galaxies: active - galaxies: formation - - galaxies: evolution - quasars: general
- black hole physics -BL Lacertae objects}

\section{Introduction}

Active galactic nuclei (AGNs) and inactive galaxies seem to be closely connected. Virtually
all local massive galaxies have central black holes (Kormendy \& Gebhardt 2001),
suggesting AGNs are a transient phase in the evolution of normal galaxies
(Cavaliere \& Padovani 1989). 
This ``Grand Unification" hypothesis is supported by several lines
of evidence.
First, the peak of black hole mass growth,
which manifests the maximum AGN activity at $z \sim 2$, is similar to the peak of the
star formation rate (e.g., Dunlop 1999; Miyaji, Hashinger, \& Schmidt et al. 2000, Wolf et al. 2003a,
Wolf et al. 2003b).
In individual AGNs, starburst activity is not uncommon,
and may be more prevalent than in normal galaxies
(Tadhunter, Dickson \& Shaw 1996; Melnick, Gopal-Krishna \& Terlevich 1997;
Aretxaga et al. 2001).
This apparent association of star formation and AGN activity
suggests that the formation of the galaxy bulge could simultaneously
trigger star formation and significant accretion onto a seed black hole.

Second, the total local black hole mass, estimated from inactive black holes
at the center of the present-day galaxies, is comparable
to the black hole mass integrated over the Hubble time
using the quasar luminosity function (Yu \& Tremaine 2002), 
indicating that dormant black holes in the present-day Universe are plausibly the end-products
of quasar activity (see also Cavaliere \& Padovani 1989).

Third, black hole mass appears to correlate with galaxy mass in a sample of $\sim 40$ nearby,
well-studied galaxies (Ferrarese \& Merritt 2000; Gebhardt et al. 2000a), with black hole mass measured from
spatially resolved kinematics.
This relation, between black hole mass ($M_{\bullet}$) and bulge stellar velocity dispersion ($\sigma$),
indicates that somehow the small-scale black hole knows the large-scale galaxy
properties, and vice versa, further supporting a physical connection between galaxy formation
and black hole accretion. 

Finally, AGN host galaxies show the same  $M_{\bullet}$-$\sigma$ correlation
(Gebhardt et al. 2000b; Ferrarese et al. 2001).
In this case, black hole masses are derived from reverberation mapping results,
which gives one of the most reliable AGN black hole mass measurements (Peterson 1993),
suggesting the overall formation processes for black holes and galaxies
are the same for normal and AGN host galaxies.

Accordingly, we started a program to test the Grand Unification hypothesis, 
by studying AGN host galaxies directly. Specifically, we investigate the Fundamental Plane of
AGN host galaxies, to see whether they obey the same tight correlation among
half light radius ($r_{e}$), surface brightness ($\mu_{e}$), and central velocity
dispersion ($\sigma$) as normal galaxies.
Already it is known that early-type AGN host galaxies, especially radio-loud AGN hosts,
follow the Kormendy relation, which is a projection onto
the $r_{e}$ -- $\mu_{e}$ plane
(Taylor et al. 1996; Urry et al. 2000; O'Dowd, Urry \& Scarpa 2002; Dunlop et al. 2003).
However, this two-parameter relation has larger scatter
than the Fundamental Plane (Dressler et al. 1987; Djorgovski \& Davis 1987; Faber et al. 1987),
and lacks the third dimension, namely, velocity dispersion.
The empirical scaling relation, among  $r_{e}$, $\mu_{e}$, and $\sigma$ observed in normal galaxies
is interpreted as evidence that 
early-type galaxies form a homologous family with a power-law relation
between galaxy mass and the mass-to-light ratio.
Due to the very low internal scatter, the Fundamental Plane is often
used to study galaxy formation (Franx 1993; Jorgensen et al. 1996; van Dokkum \& Stanford 2003), and it is a much more direct probe of the
similarities and differences between normal and AGN host galaxies.
Thus, we initiated a program to measure $\sigma$ in the host galaxies of AGN.

Ultimately the Fundamental Plane must be probed at moderate to high redshift,
where AGN are more numerous. Here we push to $z \sim 0.3$, roughly three times as high as most
previously published measurements (Bettoni et al. 2001; Barth et al. 2002; Falome et al. 2003).
Specifically, we report observations and velocity dispersion measurements
for 15 AGN host galaxies from $z = 0.05$ to $z=0.34$.
Combining with 11 published dispersion measurements,
we investigate whether early-type AGN host galaxies lie on the
same Fundamental Plane as normal galaxies, and how the mass-to-light ratio
of AGN hosts evolves relative to that of normal galaxies. 
In \S~2, we describe the observations and velocity dispersion measurements, and
in \S~3, the Fundamental Plane and its evolution.
In \S~4, we discuss black hole masses, and
in \S~5 we present the discussion and conclusions.
The correlations of black hole mass with other AGN properties will be discussed
in detail in a subsequent paper (Woo et al. 2004).
We adopt a cosmology with $\Omega=0.3$, $\Lambda=0.7$, and $H_{o}=70$ km sec$^{-1}$ Mpc$^{-1}$. 

\section{OBSERVATIONS and DATA REDUCTION}

\subsection{Sample Selection and Observation}

AGN host galaxies are more difficult to study than normal galaxies
due to the presence of bright central sources.
To determine the structural parameters, $r_{e}$, and $\mu_{e}$, 
it is necessary to subtract the AGN contribution from
a host galaxy, which then can be analyzed as a normal galaxy.
Given the importance of high spatial resolution, our first sample selection criterion was to choose
HST-imaged AGN host galaxies, for which reliable structural parameters are
already available.

Second, we chose low ratios for the AGN-to-galaxy luminosity.
Stellar absorption lines are much harder to detect from a typical bright
AGN spectrum because the brighter nuclear continuum hides stellar lines and increases
noise, and AGN broad emission lines hamper the determination
of the velocity dispersion from stellar absorption lines.
Indeed, we initially observed several typical bright AGNs with faint
host galaxies, but failed to get reliable stellar signatures even with very
high S/N ratios in the observed spectra. We concluded that it is not possible to measure the
stellar velocity dispersion for host galaxies of bright AGNs using the present
observational techniques (see also Barth et al. 2002).
Therefore, we chose BL Lac objects and radio galaxies, which typically do not have strong
emission lines, and whose luminosity is comparable to or smaller than the galaxy light.

Third, we selected low to intermediate redshift ($z \lesssim 0.4$) objects.
With 4-6m class telescopes and typical exposure times of a couple of 
hours, $z \sim 0.4-0.5$ is the approximate upper limit for measuring bulge stellar
velocity dispersion in AGN host galaxies.

Starting with the sample of 110 BL Lac objects observed with HST (Urry et al. 2000),
we selected early-type host galaxies at $z<0.4$ giving $\sim 30$ BL Lac hosts
in the southern and northern hemispheres.
In addition, we selected several well-studied radio galaxies at similar redshifts,
which were also previously observed with HST.

The spectra were obtained with the 6.5-m Magellan Clay Telescope
at Las Campanas Observatory, the CTIO 4-m telescope, and the KPNO 4-m telescope
during February to October 2003.
Table 1 shows the details of instrumental setups and the journal of observations.
We used the B\&C long-slit spectrograph at the Magellan Telescope and
the RC long-slit spectrograph at the CTIO and KPNO telescopes.
The instrumental setups were chosen to cover strong stellar
absorption lines, such as G-band (4300 \AA), Mgb triplet (around 5172 \AA),
Ca+Fe (around 5269 \AA), and NaD lines (5892 \AA) depending on the redshift,
and to provide sufficient instrumental resolution,
$\lesssim 100$ km sec$^{-1}$ (Gaussian dispersion), so that the expected
velocity dispersions ranging 150 to 350 km sec$^{-1}$
for our target galaxies could be reliably recovered.

Sky conditions were mostly photometric with excellent seeing,
with the exception of a few nights when intermittent thin clouds were present.
We used a $1^{\prime\prime}$ slit width for most of the runs. For the KPNO run we used
a $1.2^{\prime\prime}$ slit width due to worse seeing conditions.

Bias frames and flat fields were taken in the afternoon.
When the grating angle had to be changed in order to cover
different wavelength ranges, all calibration frames were
also taken again. Arc lamp spectra were taken in the afternoon for
an initial wavelength solution, and at each telescope position during the night 
before and/or after target exposures, in order to
check any systematic changes.

Late type (G8-K5) giant stars with low radial velocity ($ < 20$ km/sec)
were selected as template stars. These stars were
observed for each instrument setup, mostly at twilight.
Smaller slit widths were used for template star observations
in order to obtain a sufficiently high spectral resolution
so that the Gaussian velocity dispersions of galaxies and the redshifted templates
are comparable.

Total exposure times of 1 to 3.5 hours were used depending on
the host galaxy magnitude. Each total exposure time was typically
divided into 30 minute exposures to remove cosmic rays and avoid
CCD saturation in the presence of bright nuclei.

\subsection{Data Reduction and Dispersion Measurement}

The standard data reduction procedures, such as bias subtraction, flat-fielding,
spectral extraction, and wavelength calibration, were performed with IRAF routines.
A one-dimensional spectrum was extracted from each exposure,
using aperture sizes similar to or smaller than the effective radius
of each host galaxy (Table 2). Extracted spectra were combined to make
the final spectrum for each galaxy.
Out of 18 observed AGN hosts, 3 galaxies were removed from further analysis
since the signal-to-noise ratios of galaxy absorption lines were not sufficient to measure
the velocity dispersion.
The signal-to-noise ratio of the final sample ranges from 25 to $\sim$ 400 per pixel 
(see Table 1 for the plate scale),
high enough to measure stellar velocity dispersions.

We used a direct fitting method, in which the observed spectrum
is directly fitted in pixel space with broadened template spectra.
The best-fitting dispersion values are
determined by the smallest $\chi^{2}$ number.
This method is widely used 
(van der Marel 1994; Rix et al. 1995; Kelson et al. 2000a; Barth et al. 2002),
and has an advantage of easily removing bad pixels and AGN emission
lines by masking out specific spectral regions in pixel space. 

For the fitting procedure featureless continua with various
slopes were added to the template star spectra to account for the possible
presence of an AGN non-stellar continuum, which changes the
line strength as a function of wavelength. The observed galaxy and
template star spectra were then normalised by divinding by a continuum fit
and re-binned logarithmically.
Thus, a set of template spectra with various slopes were prepared for each template star.
Finally the template spectra were convolved with Gaussian velocity profiles with
50-400 km sec$^{-1}$, and fitted to the normalized galaxy spectrum using the
Gauss-Hermite Pixel Fitting software\footnote{available at http://www.stsci.edu/$\sim$marel/software.html},
which is described in Appendix A of van der Marel (1994).

The fitting software uses various polynomial orders
and line strength parameters to match galaxy spectra,
and determines the best $\chi^{2}$ fit, which gives
the velocity dispersion measurement.
Bad pixels (especially for CTIO 4m data), galactic absorption lines,
and various emission lines (e.g., clearly present H${\beta}$ and [OIII] lines)
were masked out before fitting.

Extensive and careful fitting with various spectral regions,
such as G-band, H${\beta}$, Mgb triplet, and NaD line regions,
were performed to determine the best-fitting spectral range.
We used the G-band region for many galaxies, and a larger spectral range
containing from the G-band to the Ca+Fe lines was chosen for 6 galaxies.
Various polynomial orders and continuum slopes were
used for fitting to test any dependence on the parameters.
Deviations with these parameters are similar to typical measurement errors.
Changing the spectral type of the template stars gives a larger variation in the velocity
dispersion. After fitting with each individual and combined template spectrum,
we chose the best-template star with the smallest $\chi^{2}$ for each galaxy.
Figure 1 shows the host galaxy spectra with the best-fitting template.

Like Barth et al. (2002), we found that Mgb triplet cannot be fit together with nearby lines
such as Fe 5270 and Fe 5335 blends. 
The Mgb triplet is much deeper in the galaxy spectrum than in the stellar template.
However, when we exclude the
Mgb triplet, the overall fit was very good. Thus, in those cases we masked out
the Mgb triplet for the final fitting (see Fig. 1a). The Mgb triplet mismatch is partially
caused by a difference in $\alpha$-enhancement between stars in massive galaxies
and Galactic stars (Barth et al. 2002). We note that the [N I] emission line, which 
in some case, is clearly present on the red side of the Mgb triplet,
will also hamper a good fit around the Mgb triplet region. 
The issue of whether more accurate values of the velocity dispersion are
obtained when the Mgb triplet lines are included or excluded was 
addressed in detail by Barth et al. (2002). They found that more
accurate results are obtained when the lines are excluded, and we 
have followed this advice. In principle it would be interesting to 
provide an independent verification of this finding, and to analyze 
in more detail what aspects of the stellar and abundance composition 
of the galaxies are the cause of this. However, this is beyond the 
scope of the present paper. It would also require a larger sample of 
higher S/N spectra.

The measured velocity dispersions ($\sigma_{m}$) are corrected for differences in
instrumental resolution between the galaxy and template spectra
using (van der Marel et al. 1997):

\begin{equation}
(\sigma_{vel})^{2} = (\sigma_{m})^{2} - [(\sigma_{inst,G})^{2}
- (\sigma_{inst,T})^{2}~].
\end{equation}
Here, $\sigma_{m}$ is the measured velocity dispersion from the fitting, and
$\sigma_{inst,G}$ and $\sigma_{inst,T}$ are the instrumental resolutions
of galaxy and template spectra, respectively.
Note that the instrumental resolution in km sec$^{-1}$ unit is independent of redshift.

For nearby galaxies, this correction is unnecessary because the instrumental
setup of template and galaxy spectra are the same.
However, this is not the case for high-redshfit galaxies because template
stars and galaxies are observed at different spectral ranges, and thus, the instrumental
resolution of the template stars should be matched to that of galaxy spectrum
(van Dokkum et al. 1996; Kelson et al. 2000a).

Figure 2 shows an example of instrumental resolution as a function of wavelength
for each instrumental setup of our Magellan run.
The resolution in \AA~ of the template spectrum is $10 - 20 \%$ better than that of the galaxy spectrum
because of the smaller slit width ($0.7^{\prime\prime}$). 
Since the template spectrum resolution is very similar to that of the galaxy spectrum 
when the template spectrum is redshifted for fitting,
the correction for the difference in instrumental resolutions is very small.
We derived instrumental resolutions at the center of each fitting spectral range for each galaxy
and corrected the galaxy velocity dispersions with Equation (1).
Typical correction values are a few km/sec.
We report corrected velocity dispersions and aperture radii for 15 AGN host galaxies
in Table 2. 

The velocity dispersion of AP Lib (1514-241) was measured previously by two groups,
Barth et al. (2002) and Falomo et al. (2003),
who reported dispersions $196 \pm 12$ km sec$^{-1}$ and $250 \pm 12$ km sec$^{-1}$, respectively.
Several other BL Lac objects have large discrepancies between
the measurements of those two groups as well.
The source of the large (4$\sigma$) discrepancy is unclear.
Our measurement for AP Lib, $\sigma = 242 \pm 7$ km sec$^{-1}$, agrees very well with 
the value of Falomo et al. (2003).

\section{The Fundamental Plane of AGN Host Galaxies}


Since the Fundamental Plane is well established among early-type galaxies
in nearby clusters (e.g., Jorgensen et al. 1996), it has become a tool   
to study the formation and evolution of normal early-type galaxies.
AGN host galaxies, at least at very low redshifts, also show the same scaling relation.
Bettoni et al. (2001) measured and collected data for 72 radio galaxies and showed that
these low redshift ($<z> \sim 0.04$) galaxies lie on the Fundamental Plane.
Barth et al. (2002) and Falomo et al. (2003) measured
bulge stellar velocity dispersions for a sample of $\sim 10$ low redshift
($<z> \sim 0.06$) BL Lac objects and showed they lie on
the same Fundamental Plane as normal galaxies.
Here, we compare the Fundamental Plane of AGN host galaxies in our 
higher redshift sample with that of low-redshift radio and normal galaxies.

We measured and collected dynamical and structural parameters for a sample
of 26 AGN host galaxies (23 BL Lac objects and 3 radio galaxies),
combining our new higher redshift data with 11 published dispersion measurements 
from Barth et al. (2002) and Falomo et al. (2003).
The velocity dispersions were corrected to a common aperture equivalent to $3.4^{\prime\prime}$ at the distance
of the Coma cluster ($z=0.023$), based on the empirical prescription of Jorgensen et al.
(1995).
For several common objects in Barth et al. (2002) and Falomo et al. (2003),
we took the mean of the logarithms of the two dispersions. The errors in the mean was calculated
by propagation of the errors in the individual measurements (see Table 3).
Note that this might underestimate the true errors, given that some of the values
of Barth et al.~(2002) and Falomo et al.~(2003) are not mutually consistent within the errors.
However, this does not affect the results of our further analysis.

For this sample of 26 galaxies,  structural parameters are collected
from the literature. The $r_{e}$ and total galaxy magnitudes are
taken from the HST snapshot survey of BL Lac objects (Urry et al. 2000),
with a few exceptions in which the HST-measured host galaxy
size ($r_{e}$) is more than factor of 2 smaller than the
ground-based result. 
These 4 cases, namely Mrk 180, Mrk 421, Mrk 501, and 3C 371, are all at very
low redshift ($z \lesssim 0.05$) and have large effective radii, $r_{e} \gtrsim 10^{\prime\prime}$.
Due to the small field of view of WFPC2, overestimated sky background would
lead to an underestimate of $r_{e}$ (Barth et al. 2002; Falomo et al. 2003).
Thus, for these four cases, we took the median of all available $r_{e}$ data
from Falomo et al. (2003).
The mean surface brightness within the $r_{e}$ is derived from the
total magnitude of the host galaxy using:

\begin{equation}
<\mu_{e}> = m_{t} + 5 log(r_e) + 2.5 log(2\pi) - A_R - K -2.5 log (1+z)^{4} ~.
\end{equation}
Here, $m_{t}$ is the total host galaxy magnitude measured in the Cousins R band 
(Urry et al. 2000),\footnote{Note that the published surface brightness in Table 2 of Urry et al. (2000) is not
extinction corrected.}
$r_e$ is the effective radius in arcsec,
$A_R$ is extinction in the R band taken from the NED database\footnote{
The NASA/IPAC Extragalactic Database (NED) is operated by the Jet
Propulsion Laboratory, California Institute of Technology,
under contract with the National Aeronautics and Space Administration.} (Schlegel et al. 1998),
and K is the K correction value interpolated from Poggianti et al.
(1997).

\subsection{Comparing with Radio and Normal Galaxies}

Radio galaxies and BL Lac objects are believed to constitute the same AGN family with different
orientation angles (Urry \& Padovani 1995).
Here we compare the Fundamental Plane of the Bettoni sample of radio galaxies along
with our AGN sample in Figure 3. 
The two panels correspond to two different edge-on projections.
The axes of the two panels are such that the 72 radio galaxies of Bettoni et al. (2001)
are viewed edge-on.
The dashed line is the Fundamental Plane fit defined for 72 radio
galaxies ({\it small dots}) in Bettoni et al. (2001), who demonstrated that
radio galaxies lie on the same Fundamental Plane as normal early-type
galaxies although the actual data points (small dots) present significant scatter (0.19 in $log~r_{e}$),
probably due to the data quality and uncertainties in color transformation.
The twenty-three BL Lac hosts and 3 radio galaxies in our sample
also follow the same correlation. 
All scales (in this figure only) are corrected to $H_{o}=50$ km sec$^{-1}$ Mpc$^{-1}$, and
$q_{o}=0$ for consistency with Bettoni et al. (2001).
Our dispersion measurements ({\it filled circles})
show a similar trend to that observed by Barth et al. (2002) and
Falomo et al. (2003; {\it open circles}).
We do not find any significant difference in the Fundamental Plane
between BL Lac hosts and radio galaxies.
The RMS scatter in our sample is 0.103 in $log~r_{e}$. Substracting the
measurement errors in quadrature, we find the intrinsic scatter to be
0.071 in  $log~r_{e}$.

We also compare the normal galaxy Fundamental Plane with that of AGN host galaxies
to test the ``Grand Unification'' picture.
Figure 4 shows two edge-on projections of the Fundamental Plane.
The axes of the two panels are such that the Coma cluster galaxies 
in Jorgensen et al.~(1996) are viewed edge-on.
The dashed line is a fit to the Coma cluster galaxies (small dots) from Jorgensen et al. (1996).
Our sample of AGN host galaxies, converted from the Cousins R band to the Gunn r band
assuming r-R=0.35 (Jorgensen 1994) are plotted with different symbols based on redshift. 
This figure shows that AGN host galaxies lie on the same Fundamental
Plane as normal galaxies. The higher redshift host galaxies lie below the dashed line
because the surface brightness is brighter on average due to the
younger stellar populations, assuming $r_{e}$ and $\sigma$ are not changed.
This trend is also seen in cluster and field early-type galaxies in
similar redshift ranges (van Dokkum et al. 1996; Treu et al. 2001),
and corresponds to an evolution of the Fundamental Plane.
By correcting passive luminosity evolution with a stellar population synthesis model with
formation redshift $z_{form} = 2$,
the Fundamental Plane becomes tighter (see \S~3.2).

\subsection{Mass-to-light ratio evolution}

The mass-to-light ratio is a fundamental property of galaxies
and a basic tool for studying galaxy evolution (Franx 1993).
If early-type galaxies formed at relatively high redshift and 
evolved passively afterward, the evolution of the Fundamental Plane is simply caused by
the dimming of surface brightness due to the aging of the stellar population
(Tinsley \& Gunn 1976). 
However, if early-type galaxies experience mergers at relatively low
redshift, the evolution of the Fundamental Plane will be different due to the ensuing
star formation.
Thus, the study of the mass-to-light ratio evolution through the Fundamental
Plane can constrain the galaxy formation history.

Various recent studies of high redshift (up to $z \sim 1$) early-type galaxies have 
attempted to test galaxy formation scenarios (e.g., monolithic collapse versus merging models) 
by estimating the star formation epoch,
showing that
field and cluster early-type galaxies formed at $z \gtrsim 1$ and evolved passively afterward
(Kelson et al. 2000b; Treu et al. 2001; van Dokkum et al. 2001; Treu et al. 2002; 
Gebhardt et al. 2003; Rusin et al. 2003; van Dokkum \& Stanford 2003; van der Wel et al. 2004).
The evolution of the mass-to-light ratio in early-type cluster galaxies ($ 0.02 < z < 1.27$) 
is consistent with a single burst of star formation at $z \sim 2$--3, but we note that at higher
redshift the picture may be more complex. One of 3 galaxies at $z =1.27$ shows much lower mass-to-light ratio and
strong Balmer absorption lines, indicating a younger stellar population than predicted
with $z_{form}=2$--3 models (van Dokkum \& Stanford 2003). 

In the case of early-type field galaxies, the formation redshift ranges from $z \sim 1$--3,
with at least some observational studies reporting that early-type field galaxies
show a steeper mass-to-light ratio evolution (van Dokkum et al. 2001; Treu et al. 2002; Gebhardt et al. 2003).
This suggests the average age of their stellar population is younger than that of
cluster galaxies, qualitatively consistent with hierarchical merging models that predict
early-types in the field formed later than in clusters due to the less dense environment 
(Kauffman 1996; Diaferio et al. 2001). 
However, the age difference between field and cluster galaxies, estimated from the observed
mass-to-light ratios, is much smaller than predicted with hierarchical merging models (van Dokkum
et al. 2001; Treu 2004).
It is not even clear whether field galaxies formed later than cluster galaxies,
since other studies suggest that field and
cluster early-type galaxies have a comparable formation epoch with similar stellar population ages
(van Dokkum \& Ellis 2003; Rusin et al. 2003).
At each epoch of observation there are plausibly various early-type galaxies
with stellar populations formed at different epochs ($z_{form}=1$---3),
since there is a large scatter in the mass-to-light ratio
among field early-type galaxies up to $z \sim 1$, indicating a significant
age spread among field galaxies (van de Ven et al. 2003; van der Wel et al. 2004).

If AGN host galaxies are fundamentally normal galaxies with an active nucleus,
the evolution of the mass-to-light ratio should be similar to that of normal
galaxies. In this section, we investigate the mass-to-light ratio evolution of 
our sample of 26 AGN host galaxies.

Because of the virial theorem, galaxy masses can be derived with
\begin{equation}
M = K~ \sigma^{2}~r_{e}~ /~ G,
\end{equation}
where $K$ is a constant and $r_{e}$ is the effective radius. 
By definition, the luminosity of a galaxy is given by $L = 2 \pi r_{e}^{2} I_{e}$,
where $I_{e}$ is the average surface brightness within the effective radius. 
Hence, the mass-to-light ratio is given by
\begin{equation}
log~ M/L = 2~ log~\sigma - log~ I_{e} - log~ r_{e} +C,
\end{equation}
where,  the constant $C = log (K/ 2 \pi G)$. 
In the Gunn r band, $log~I_{e} \equiv -0.4 (<\mu_{e}> - 26.4)$,
where $I_{e}$ is in $L_\odot pc^{-2}$ and $\mu_{e}$ is 
the surface brightness in $mag~ arcsec^{-2}$ (Jorgensen et al. 1996).
If galaxies evolve without merging or accretion,
the mass-to-light ratio evolves due to the luminosity evolution of the stellar populations 
while the mass, $\sigma$, and $r_{e}$ remain the same (van Dokkum \& Franx 1996).
Using these relations, we estimated the mass-to-light ratio for each host galaxy, 
and its evolution by normalizing it with the Fundamental Plane of the Coma cluster galaxies.
All parameters were corrected to a common set of cosmological parameters, 
$H_{o}=70$ km sec$^{-1}$ Mpc$^{-1}$, $\Omega=0.3$, and $\Lambda=0.7$.

We also modeled the evolution of the mass-to-light ratio in the Gunn r band
using stellar population synthesis models with the Salpeter IMF (Bruzual \& Charlot 2003).
The star formation rate is assumed to be exponentially decreasing: 
\begin{equation}
SFR \propto exp^{-t~/~\tau},
\end{equation}
where {\it t} is the time elapsed since the galaxy formed and
$\tau$ is the e-folding time of the star formation rate.

Figure 5 shows the passive evolution model predictions of the mass-to-light ratio
in the Gunn r band, depending on the formation redshift and $\tau$.
To address the evolution of AGN hosts we compare high-redshift hosts to low-redshift hosts.
We define $\Delta \log (M/L)$ to be difference in logarithm between the $M/L$ at a
certain redshift and the $M/L$ in our lowest AGN host redshift bin,
which is centered on $z=0.046$.
The Figure shows both the mass-to-light ratio of individual host galaxies (open circles) and
the averaged $M/L$ values over each redshift bin (filled circles). 
There is a clear $M/L$ decrease as a function of redshift. 
The mass-to-light ratio at $z=0.046$ corresponds to a $38\%$ increase since $z=0.324$,
which can be reproduced by a single burst model with $z_{form} \sim 2$,
although the data points are consistent with the passive evolution models of 
$z_{form} \gtrsim 1$ within the 68 $\%$ confidence limit based on $\chi^{2}$ analysis.
The mass-to-light ratio evolution in the Gunn r band corresponds to
$\Delta log (M/L_{r})/ \Delta z = -0.51 \pm 0.18$. Using the stellar population
models consistent with this value, we estimated the $M/L$ 
evolution in the rest-frame B band, $\Delta log (M/L_{B})/ \Delta z \sim -0.62$.
The trend of the $M/L$ evolution of our AGN host galaxies is similar to that of normal galaxies
in the field and clusters with star formation epoch at $z_{form}=1$---3 
($\Delta log (M/L_{B})/ \Delta z = 0.46$--0.72; Treu et al. 2002; Rusin et al. 2003;
van de Ven et al. 2003; van Dokkum \& Stanford 2003).
A larger sample of higher redshift ($z >0.3$) galaxies is clearly required to narrow down
the formation epoch of AGN host galaxies (O'Dowd \& Urry 2004). 
We note, however, that 
a higher $z_{form}$ model with extended star formation (thin-solid line)
and a lower $z_{form}$ single burst model (dashed line) predict the very similar mass-to-light ratio evolution, especially at low redshift.

Correcting the surface brightness of the galaxies in our sample using the single burst model (with $z_{form}=2$)
predictions for the $M/L$ evolution, as in Figure 4, 
the Fundamental Plane shows a tight correlation within RMS scatter 0.095 in 
$log~r_{e}$. Subtracting the measurement errors in quadrature, we find the intrinsic
scatter of the Fundamental Plane to be 0.080 in $log~r_{e}$, 
which is close to the intrinsic scatter, 0.070, of the Fundamental Plane of
nearby clusters (Jorgensen et al. 1996).

We also estimated each galaxy mass using Equation (3), with $K=5$ as suggested by Bender
et al. (1992).
AGN host galaxies show the same trend as normal galaxies in
that more massive galaxies have larger mass-to-light ratios (Fig. 6).
Using the luminosity correction with $z_{form}=2$ model, we obtain
$M/L \propto M^{0.38 \pm 0.06}$, which is similar to that of normal galaxies (Jorgensen et al. 1996)

We conclude that there is no significant difference of the $M/L$ evolution 
between normal and AGN host galaxies. The AGN host galaxies show the same Fundamental
Plane, the same $M/L$ relation, and a similar star formation epoch.

\section{AGN BLACK HOLE MASS}

\subsection{Black Hole Mass of BL Lac Objects and Radio Galaxies}

AGN black hole masses estimated from reverberation mapping results
show the same correlation with bulge kinematics as normal galaxies
(Gebhardt et al. 2000b; Ferrarese et al. 2001).
This tight correlation between black hole mass and bulge dispersion 
plausibly holds up to the epoch when galaxies have formed and virialized (Shields et al. 2003),
after which the kinematical properties of each galaxy should
be approximately stationary. Therefore, the mass-dispersion correlation
is probably valid for galaxies that lie on the Fundamental Plane.

Our sample of AGN host galaxies at $z<0.34$ shows the same Fundamental Plane
as normal and radio galaxies, particularly  at similar redshift range common
to all samples, as we shown in \S~3. Thus, we estimate the black hole mass of
these objects using the $M_{\bullet}$-$\sigma$ relation of normal galaxies (Tremaine et al. 2002):
\begin{equation}
M_{\bullet} = 1.349 \times 10^{8} M_{\odot} (\sigma_{e}/200 {\rm km~s}^{-1})^{4.02} ~.
\label{sig}
\end{equation}
with proper aperture corrections for $\sigma_{e}$, following Jorgensen et al.
(1995; which gives at most a 10 \% change in the velocity dipersion value).

In Figure 7, we show the distribution of black hole masses for
BL Lac objects, ranging from a few tens of millions
to a billion solar masses, similar to the black hole mass ranges of 
normal (Kormendy \& Gebhardt 2001) and radio galaxies (Woo \& Urry 2002). 
It seems that regardless of the strength of the AGN activity,
black hole mass spans a similar range, 
although the upper and lower limits of the black hole mass are not constrained.

The spectral energy distributions of BL Lac objects show two distinct peaks,
at low and high frequencies, due to synchrotron and inverse Compton emission from
the unresolved nucleus.
The peak frequencies vary systematically and BL Lacs have been divided into
two types, namely, low-energy peaked (LBL) and
high-energy peaked (HBL). Using measured velocity dispersions,
Barth et al. (2002) and Falomo et al. (2003) derived black hole masses using the
$M_{\bullet}$-$\sigma$ relation and found similar BH mass ranges between LBL and HBL, 
as predicted by Urry et al. (2000) on the basis of their similar host
galaxy luminosities.
With a larger sample of BL Lac objects, we confirm that 
black hole mass is independent of BL Lac type,
suggesting black hole mass does not control the peak frequency (Fig. 7). 

\subsection{Black Hole Mass to Galaxy Luminosity Relation}

We derived the black hole mass to host galaxy luminosity relation in the Cousins R band (Fig. 8).
For the 11 objects at $z <0.1$, the best fit is:

\begin{equation}
M_{\bullet}/M_{\odot} = -0.51(\pm 0.11) \times M_{R} -3.23(\pm 2.47)~.
\label{sig}
\end{equation}
Here, $M_{R}$ is the rest-frame absolute R magnitude of the host galaxy, determined using
the apparent magnitude, after correcting for foreground Galactic extinction and applying
a K correction.
Higher redshift objects are removed before fitting since younger
stellar populations can alter the local ($z < 0.1$) black hole mass -- galaxy luminosity relation. 
The derived slope of the mass -- luminosity relation in the Cousins R band is
-0.51, similar to -0.50 derived for local normal and radio galaxy samples
(McLure \& Dunlop 2002; Bettoni et al. 2003),
indicating the local mass -- luminosity relation is indistinguishable among
normal, radio, and AGN host galaxies.
The mass -- luminosity relation derived from a sample of large redshift range suffers
intrinsic scatter because host galaxy luminosity evolves as a function of time
while the galaxy mass remains the same, probably up to the epoch of galaxy
formation (see Fig. 4 of McLure \& Dunlop 2002).
Also, if derived from the local mass -- luminosity relation,
the black hole mass of high-redshift AGN will be overestimated
due to the younger stellar population.

Therefore, we looked explicitly at the effect of luminosity evolution. 
We corrected all absolute R magnitudes to $z=0$ using a stellar population synthesis model
with $z_{form}=2$. We then derived the mass -- luminosity relation 
for high- ($z>0.1$) and low-redshift ($z<0.1$) galaxies. The slope between
mass and luminosity for the high-redshift sample is $-0.45 \pm 0.12$, lower than
that of the low redshift sample ($-0.52 \pm 0.11$)  but consistent within the uncertainties.
The values of y-intercept are also consistent within the uncertainties
($-3.33\pm2.45$ for the low-redshift sample; $-2.03\pm2.73$ for the high-redshift sample).
The RMS scatter increases from 0.16 to 0.24 in $log M_{\bullet}$,
suggesting the intrinsic scatter of the relation is larger at high redshift.
However, it might be simply due to the larger luminosity correction error
for younger galaxies at higher redshift.


\section{DISCUSSION and CONCLUSION}

We report new stellar velocity dispersion measurements for 15 AGN host galaxies at $z
\lesssim 0.34$.
Combining with published measurements for 11 low redshift ($<z> \sim 0.06$) BL Lac hosts and
72 low redshift ($<z> \sim 0.04$) radio galaxies, we find that
AGN host galaxies lie on the same Fundamental Plane as normal galaxies.
The mass-to-light ratio increases $\sim 40\%$ since $z \sim 0.3$, which is
similar to that of normal early-type galaxies, and consistent with single burst
galaxy formation models with $z_{form} \gtrsim 1$,
indicating AGN host galaxies formed at high redshift and evolved passively afterward.
The fact that AGN host galaxies show the same Fundamental Plane
and mass-to-light ratio evolution as normal galaxies indicates that there is no
fundamental difference between AGN hosts and normal early-type galaxies.
Although BL Lac objects (and radio galaxies) do not represent all families of AGNs, 
this is yet another step to confirm AGN as a transient phase of normal galaxy
evolution, also known as the Grand Unification hypothesis. 
More studies, including radio-quiet AGNs and more AGNs at high redshift, are required to further verify this.
However, measurements of bulge stellar velocity dispersion for host galaxies with a bright nucleus
are extremely difficult due to the presence of luminous AGN continuum. 
New techniques will be necessary to overcome this barrier.

We estimated black hole masses for the combined sample of AGN host galaxies using the $M_{\bullet}$-$\sigma$ correlation.
The black hole mass distribution of 23 BL Lac objects is similar to that of 72 local radio galaxies.
We also confirm that black hole mass is independent of BL Lac object type,
suggesting the peak frequency of BL Lac SED is controlled by other parameters.

The correlation between black hole mass and galaxy luminosity for a sub-sample of
AGNs at $z < 0.1$ has a slope consistent with that of local radio and normal galaxies.
The mass -- luminosity relation derived for larger redshift AGN is
uncertain due to the luminosity evolution of the host galaxies.

\acknowledgements
This research is a part of the AGN key project of the Yale-Calan collaboration, and has been
supported by Fundacion Andes. Partial support was also provided by
NASA grant GO09122.09-A, GO-09121.05-A
from Space Telescope Science Institute, which is operated by the Association of
Universities for Research in Astronomy, Inc., under NASA contract NAS 5-26555.
We thank the referee for suggesting many constructive points.
J. W. thanks the Departamento de Astronomia, Universidad de Chile
for the hospitallity during the Yale-Calan collaboration meetings in Santiago.

\clearpage
\begin{figure}
\plotone{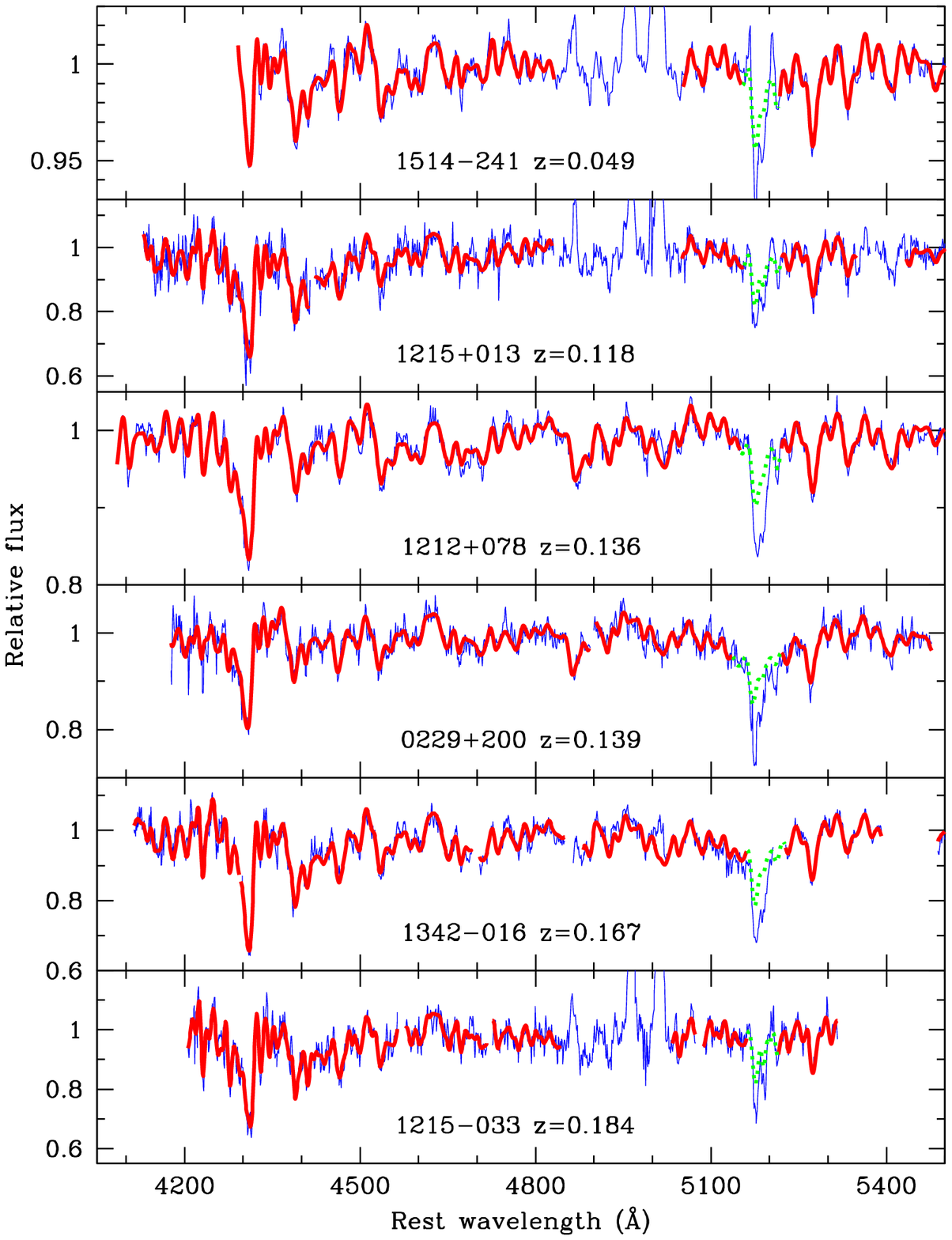}
\end{figure}

\begin{figure}
\plotone{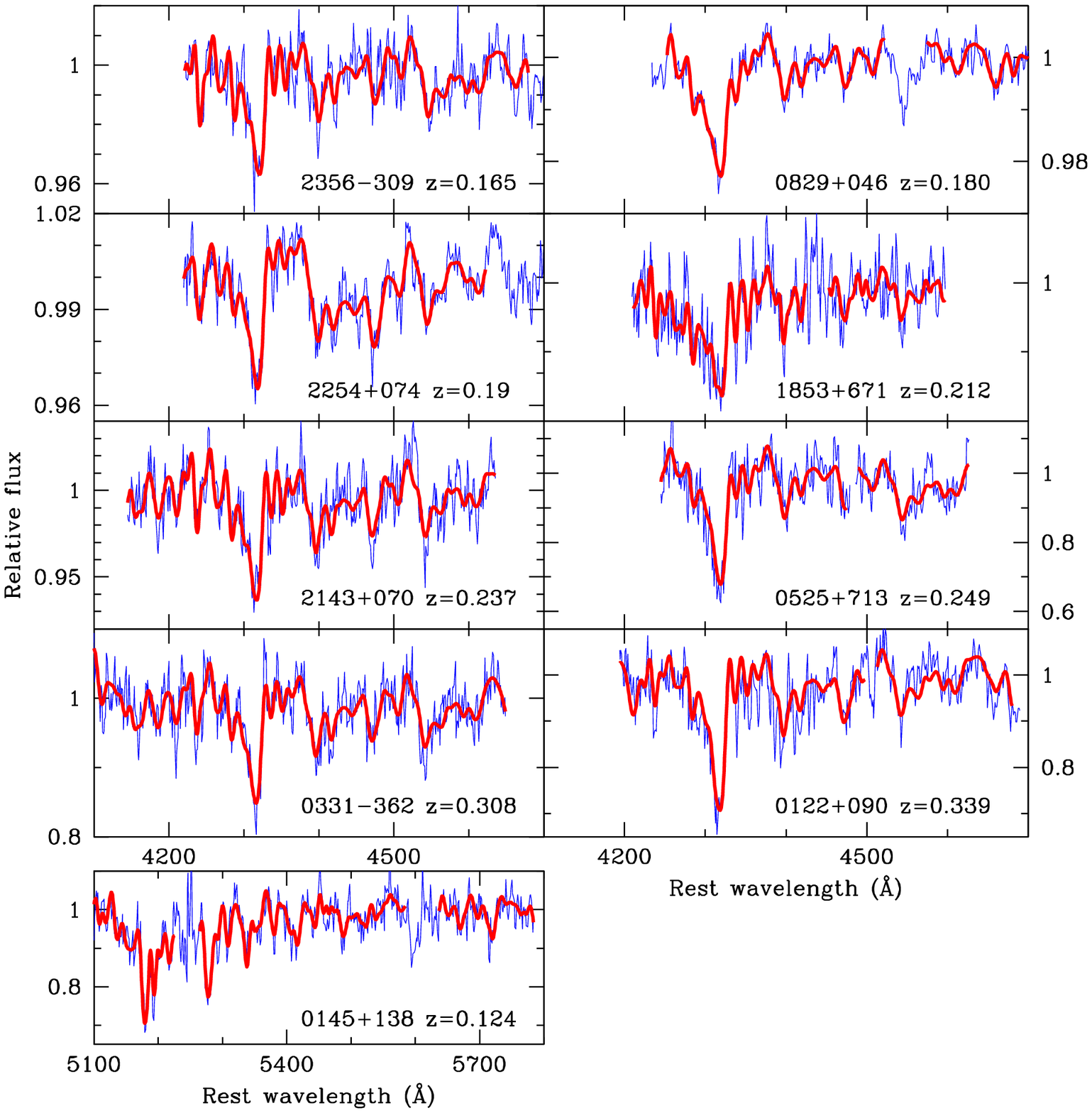}
\caption{
Observed spectra of AGN host galaxies, with best-fit template.
{\it Top panel:} Six galaxies with larger wavelength fitting range.
Major stellar absorption lines, such as the G-band (4300 \AA),
$H{\beta}$ (4861 \AA), Mgb triplet (around 5172 \AA), and Ca+Fe (around 5269 \AA)
are clearly visible in the rest-frame galaxy spectra ({\it thin line} [colored blue
in electronic edition]). 
The best-fit template spectrum broadened with a Gaussian velocity dispersion is 
over-plotted for each galaxy ({\it thick line} [colored red in electronic edition]). Bad pixels and AGN emission lines (especially
$H{\beta}$ and [O III] lines) were masked out before fitting.
The Mgb triplet line is not well matched with the template spectra and therefore has been masked
out ({\it dotted line} [colored green in electronic edition]).
{\it Bottom panel:} Nine galaxies with smaller fitting ranges.
}
\end{figure}

\begin{figure}
\plotone{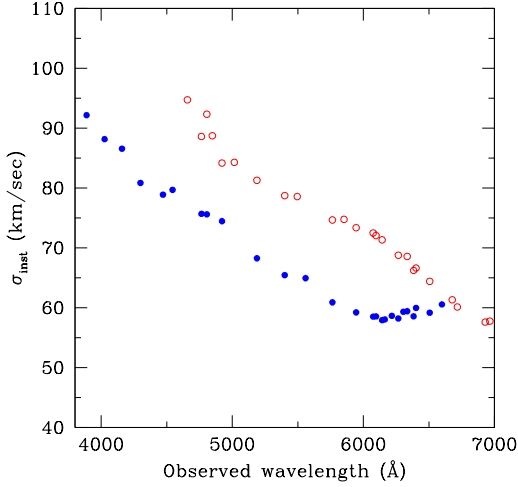}
\caption{Instrumental resolution of template ({\it filled circles}) and galaxy spectra ({\it open circles})
as a function of wavelength as obtained with the Magellan telescope. The resolution was determined
as Gaussian dispersion measurements from He-Ne-Ar Arc lamp lines for each instrumental setup.
The resolution in \AA~ of the template spectrum is $\sim 10 - 20 \%$ better due to the smaller
slit width ($0.7^{\prime\prime}$), so that the resolution of the galaxy and redshifted template
spectra are very similar over the fitted spectral range.
}
\end{figure}

\begin{figure}
\plotone{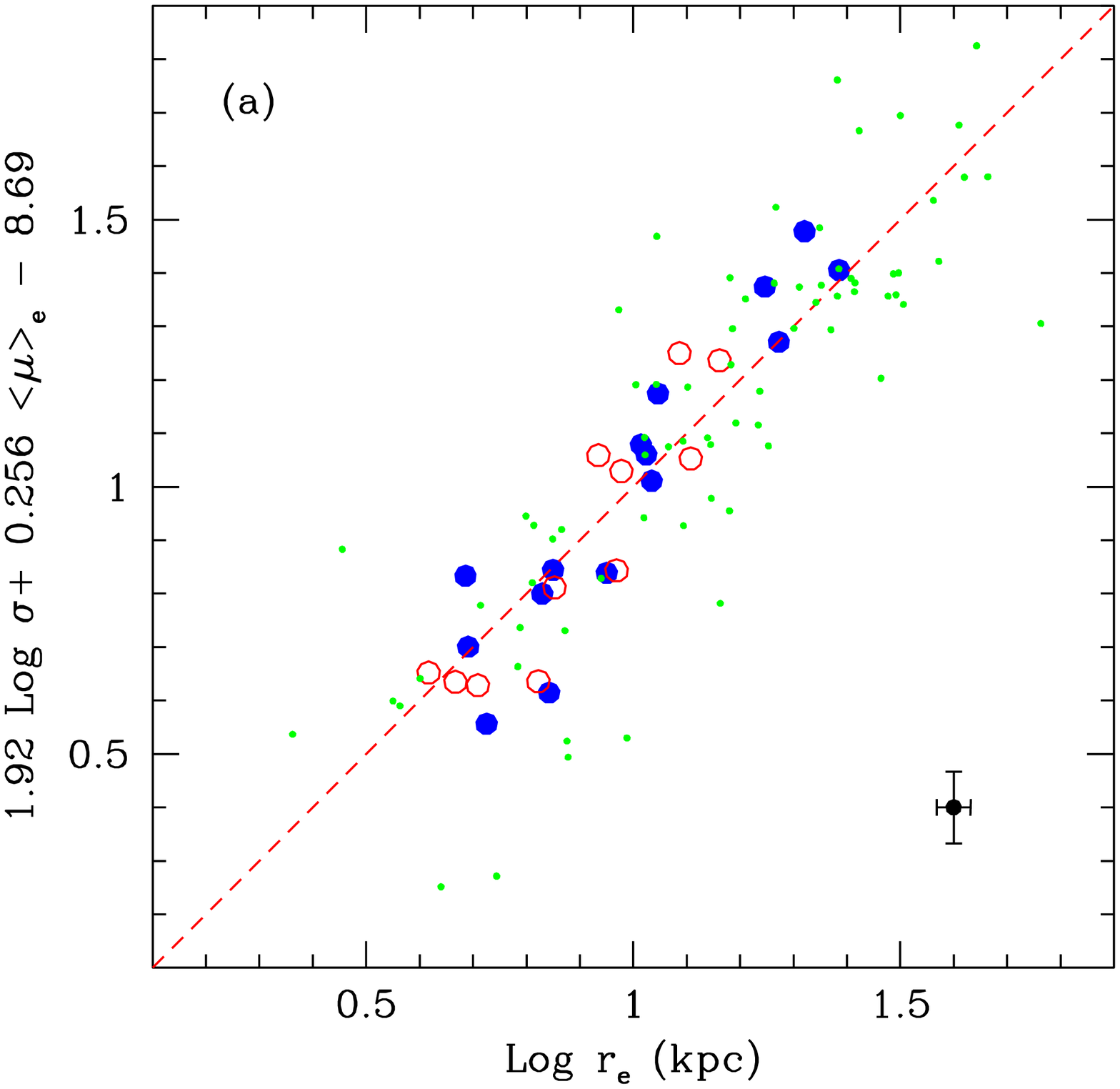}
\end{figure}

\begin{figure}
\plotone{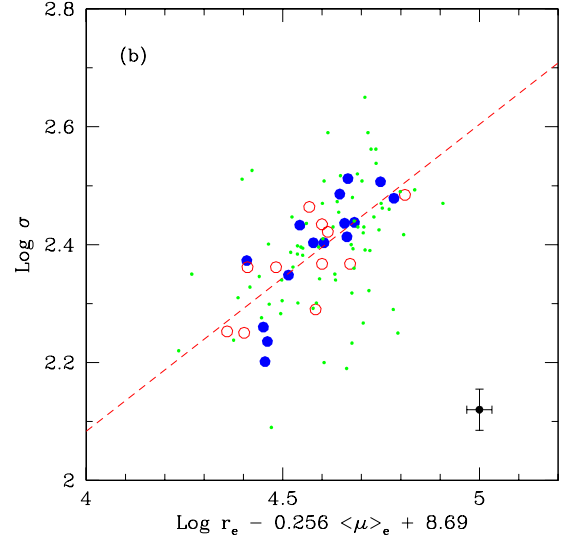}
\caption{The Fundamental Plane of AGN host galaxies. 
{\it Filled circles:} BL Lac hosts and radio galaxies from spectroscopically observed sample.
{\it Open circles:} BL Lac hosts from Barth et al. (2002) and Falomo et al. (2003).
{\it Small dots:} radio galaxies from Bettoni et al. (2001).
{\it a}) Edge-on view of the Fundamental Plane for 26 AGN hosts showing
a tight correlation among the three parameters
with RMS scatter 0.103 in $log~r_{e}$. 
The dashed line is the Fundamental Plane relation
defined by Bettoni et al. (2001), who used a 72 low redshift ($<z> \sim 0.04$)
radio galaxy sample. The error bars on the lower right represent typical
errors for our observed sample.
{\it b}) A different edge-on view of the Fundamental Plane.
}
\end{figure}

\begin{figure}
\plotone{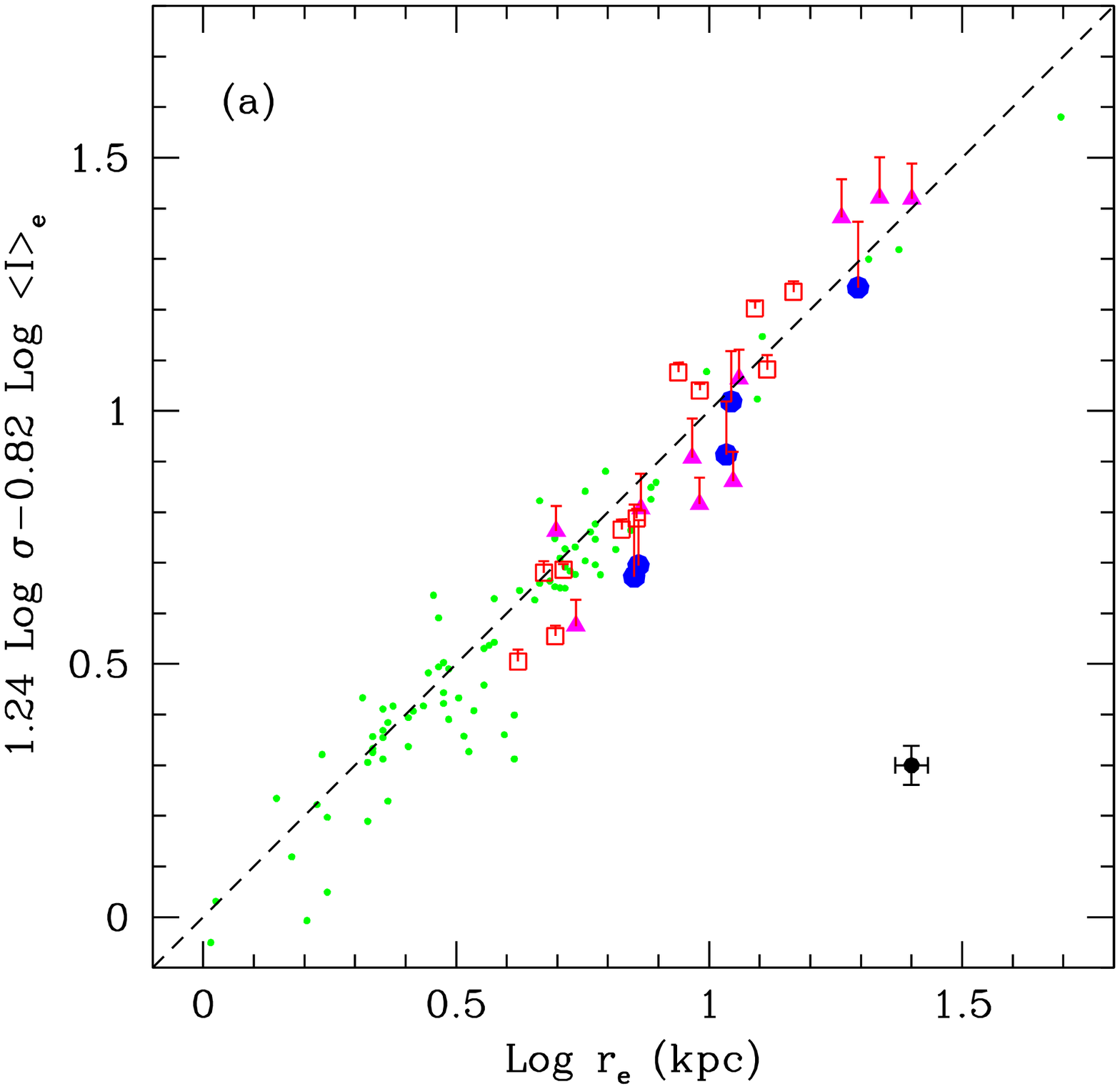}
\end{figure}
\begin{figure}
\plotone{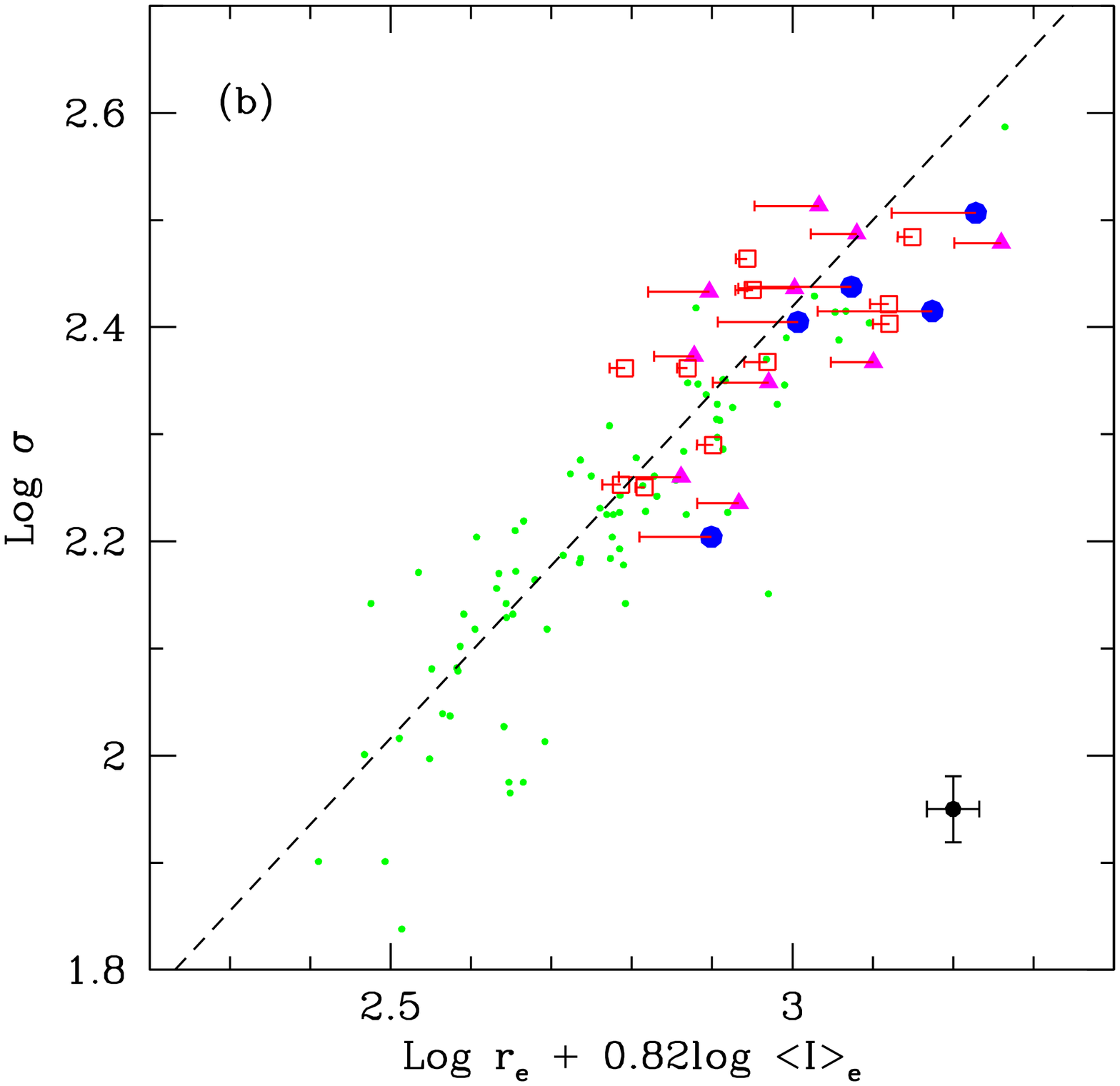}
\caption{The Fundamental Plane of AGN host and normal Galaxies, separated by redshift
interval, in two edge-on views.
{\it Squares:} AGN hosts at $z<0.1$;
{\it triangles:} AGN hosts at $0.1<z<0.2$;
{\it circles:} AGN hosts at $0.2<z$. 
Our sample of 26 AGN host galaxies lie along the same Fundamental Plane ({\it dashed line}),
defined with normal early-type galaxies in the nearby Coma cluster (Jorgensen et al. 1996; {\it small dots}).
The AGN host galaxy Cousins R bands are transformed to Gunn r, using $r-R=0.35$.
Host galaxies with higher redshift in our sample lie below the fit due to the
brighter luminosity, indicating a younger age of the stellar population.
Correcting with a passive evolution model with $z_{form}=2$ changes the points
as indicated by the short lines. The error bars on the lower right represent
typical errors for our observed sample.
}
\end{figure}

\begin{figure}
\plotone{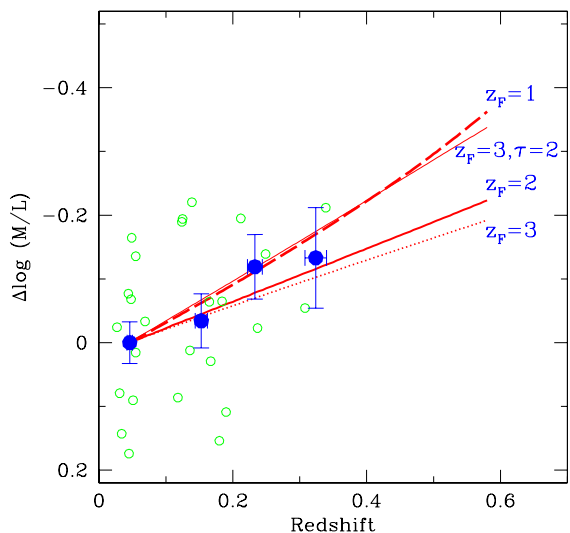}
\caption{
Evolution of the mass-to-light ratio of AGN host galaxies. 
The evolution of M/L is consistent with single burst models with $z_{form} \gtrsim 1$.
{\it Open circles:} individual host galaxies;
{\it filled circles:} averaged $\Delta log (M/L)$ for each redshift bin with 1 $\sigma$ error bars;
{\it dashed line:} stellar population synthesis model with single burst at $z_{form}=1$;
{\it solid line:} a single burst model with $z_{form}=2$;
{\it dotted line:} a single burst model with $z_{form}=3$;
{\it thin-solid line:} an extended star formation model with $z_{form}=3$ and $\tau=2$ Gyr, using Eq. (5).
The observed trend is similar to that of normal early-type galaxies (Treu et al. 2002; van Dokkum \& Stanford 2003).
}
\end{figure}

\begin{figure}
\plotone{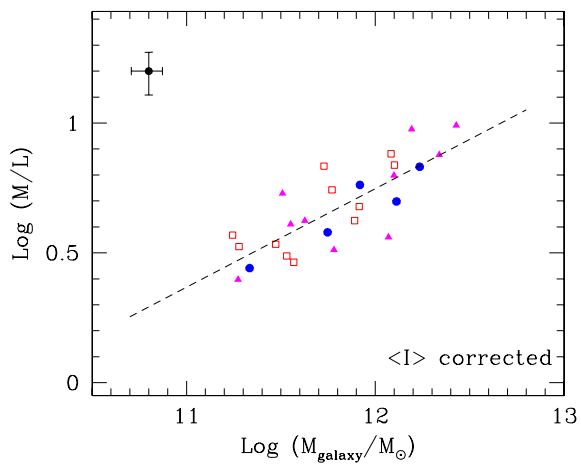}
\caption{Mass-to-light ratio as a function of galaxy mass, 
corrected for the $M/L$ evolution using the $z_{form}=2$ model.
Symbols are the same as in Fig. 4. Like normal early-type galaxies,
the 26 AGN host galaxies have a power law relation between $M/L$ and galaxy mass.
The dashed line corresponds to the best fit to the data, with a slope of 0.38, similar to
that of normal galaxies (Jorgensen et al. 1996). The error bars on the upper left represent typical errors.
}
\end{figure}

\begin{figure}
\plotone{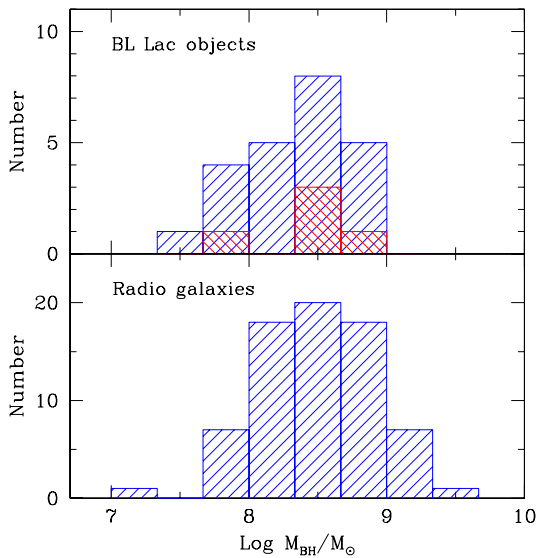}
\caption{Black hole mass distribution of BL Lac objects ({\it upper panel})
and radio galaxies ({\it lower panel}). 
Black hole masses for 72 radio galaxies are from Woo \& Urry (2002),
which were derived using the $M_{\bullet}$-$\sigma$ relation (Eq. 6) and dispersion measurements of Bettoni et al. (2001).
Black hole masses for our sample are also estimated from the host galaxy velocity dispersion
using $M_{\bullet}$-$\sigma$ relation.
The black hole mass of our 23 BL Lac sample spans a similar range as radio galaxies.
Black hole mass is independent of the BL Lac type, as shown by the
single-hatched (cross-hatched) area representing the high-frequency-peaked (low-frequency-peaked) BL Lac objects.
}
\end{figure}

\begin{figure}
\plotone{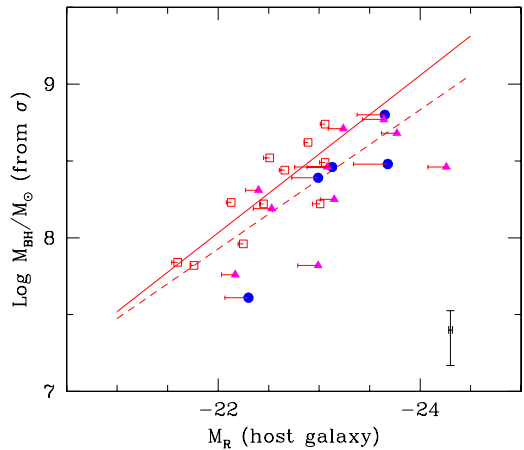}
\caption{Black hole mass to host galaxy luminosity relation in the Cousins R band.
Symbols are the same as in Fig. 4.
The local mass -- luminosity relation is derived using 11 lower redshift ($z < 0.1$) objects
({\it solid line}). 
Higher redshift host galaxies fall below this correlation due to the mass-to-light ratio evolution.
High redshift AGN black hole masses derived from the local mass -- luminosity relation
will be overestimated because of the brighter host galaxy luminosity due to the younger stellar population. The R band magnitudes corrected for luminosity evolution with a stellar population
synthesis model with $z_{form}=2$ are indicated by short lines. The mass -- luminosity relation
for higher redshift objects, corrected to $z=0$, is shown with the
dashed line. The slopes and y-intercepts of this relation for high- and
low-redshift objects are consistent 
within the uncertainties although the RMS scatter of higher redshift objects
is larger. The error bars on the lower right represent typical errors. The error 
in black hole mass is the propagated error from the measured stellar velocity dispersion.
}
\end{figure}

\clearpage
\begin{deluxetable}{lcrrrrrrrrr}
\tablewidth{0pt}
\tablecaption{Journal of observations}
\tablehead{
\colhead{Run} &
\colhead{Date} &
\colhead{Telescope} &
\colhead{Inst.} &
\colhead{l/mm} &
\colhead{slit}&
\colhead{res.} &
\colhead{\AA~pixel$^{-1}$} &
\colhead{scale} &
\colhead{seeing}  &
\colhead{sky}
 }
\startdata
1 & 2/24,25/03  & Magellan 6.5m & B\&C & 600 & 1   & 72 & 1.56 & 0.25 & 0.7-0.9 & clear  \\
2 & 4/28,29/03  & CTIO 4m       & RC   & 527 & 1   & 66 & 1.2  & 0.5  & 1.0-1.5 & thin   \\
3 & 8/27/03     & CTIO 4m       & RC   & 527 & 1   & 66 & 1.2  & 0.5  & 0.8-1.0 & clear  \\
4 & 9/15,16/03  & Magellan 6.5m & B\&C & 600 & 1   & 72 & 1.56 & 0.25 & 0.5-0.9 & clear  \\
5 & 10/4,5/03   & KPNO 4m       & RC   & 632 & 1.2 & 41 & 1.42 & 0.69 & 1.2-1.5 & clear  \\
\enddata
\label{observation}
\tablenotetext{a}{Column 
(1) observing run,
(2) observing date, 
(3) telescope,
(4) instrument,
(5) grating,
(6) slit width in arcsec,
(7) instrumental resolution (Gaussian dispersion) in km sec$^{-1}$ at 6000 \AA,
(8) plate scale in \AA~pixel$^{-1}$,
(9) spatial scale in $^{\prime\prime}$ pixel$^{-1}$,
(10) seeing in arcsec, recorded from guiding cameras,
(11) sky condition.
}
\end{deluxetable}

\begin{deluxetable}{lcrrrrrrrrrrrrrr}
\rotate {}
\tablewidth{0pt}
\tablecaption{Targets and Measurements}
\tablehead{
\colhead{Name}        &
\colhead{z}           &
\colhead{$<\mu_{e}>$} &
\colhead{$A_R$}       &
\colhead{$K_R$}       &
\colhead{$m_R$}       & 
\colhead{$\Delta m_R$}       & 
\colhead{$r_e$}       &
\colhead{$\Delta r_e$}&
\colhead{$\sigma$}    &
\colhead{$\Delta \sigma$} &
\colhead{Run}         &
\colhead{Exp.}        & 
\colhead{r}           &
\colhead{S/N}         &
\colhead{C}  \\                
\colhead{} &
\colhead{} &
\colhead{} &
\colhead{} &
\colhead{} &
\colhead{} &
\colhead{} & 
\colhead{$^{\prime\prime}$}      &
\colhead{$^{\prime\prime}$}      &
\colhead{km sec$^{-1}$}   &
\colhead{km sec$^{-1}$}   &
\colhead{}            &
\colhead{hour}        & 
\colhead{$^{\prime\prime}$}      &
\colhead{ } &
\colhead{ } \\                    
\colhead{1} &
\colhead{2} &
\colhead{3} &
\colhead{4} &
\colhead{5} &
\colhead{6} & 
\colhead{7} &
\colhead{8} &
\colhead{9} &
\colhead{10} &
\colhead{11} &
\colhead{12} & 
\colhead{13} &
\colhead{14} &
\colhead{15} &
\colhead{16} }
\tablecolumns{16}
\startdata
0122+090       &  0.339& 18.97&  0.25& 0.495 &18.88&  0.04& 1.05&  0.10&  240.&   22. & 4 & 1.7 & 1.25& 52 &  1.08\\ 
0145+138       &  0.124& 19.35&  0.16& 0.154 &16.96&  0.03& 1.75&  0.05&  155.&   21. & 5 & 2.5 & 4.8 & 32 &  1.11\\
0229+200       &  0.139& 19.30&  0.36& 0.175 &15.85&  0.01& 3.25&  0.07&  267.&   17. & 5 & 3   & 6.9 & 49 &  1.13\\
0331-362       &  0.308& 20.63&  0.04& 0.431 &17.81&  0.02& 3.10&  0.20&  249.&   20. & 4 & 2.0 & 2   & 68 &  1.10\\
0525+713       &  0.249& 19.36&  0.32& 0.327 &17.49&  0.01& 1.98&  0.01&  279.&   28. & 5 & 3.5 & 6.9 & 25 &  1.15\\
0829+046       &  0.180& 21.07&  0.09& 0.231 &16.94&  0.04& 4.30&  0.75&  250.&   37. & 1 & 3.5 & 2   & 443&  1.09\\
1212+078       &  0.136& 19.89&  0.06& 0.171 &16.02&  0.01& 3.40&  0.10&  283.&    6. & 1 & 1.5 & 2.5 & 133&  1.08\\
1215+013$^{a}$ &  0.118& 19.40&  0.06& 0.146 &16.50&  -   & 1.67&  -   &  221.&   13. & 2 & 1.5 & 2   & 41 &  1.07\\
1215-033$^{a}$ &  0.184& 20.27&  0.10& 0.236 &17.10&  -   & 2.14&  -   &  166.&   13. & 2 & 3   & 2.5 & 43 &  1.10\\
1342-016$^{a}$ &  0.167& 21.16&  0.13& 0.213 &15.60&  -   & 6.29&  -   &  248.&    9. & 2 & 3   & 3   & 66 &  1.10\\
1514-241       &  0.049& 18.66&  0.37& 0.052 &14.45&  0.01& 3.70&  0.10&  242.&    7. & 1 & 1   & 2.5 & 302&  1.05\\
1853+671       &  0.212& 19.84&  0.12& 0.274 &18.19&  0.01& 1.50&  0.08&  146.&   31. & 5 & 3   & 2   & 28 &  1.09\\
2143+070       &  0.237& 20.06&  0.20& 0.309 &17.89&  0.02& 2.10&  0.15&  232.&   19. & 4 & 1.5 & 2   & 103&  1.09\\
2254+074       &  0.190& 20.88&  0.18& 0.244 &16.61&  0.02& 4.90&  0.35&  300.&   18. & 4 & 1.5 & 2   & 207&  1.09\\
2356-309       &  0.165& 19.63&  0.04& 0.211 &17.21&  0.02& 1.85&  0.10&  206.&   16. & 4 & 1   & 2   & 151&  1.08\\
\enddata
\label{data}           
\tablenotetext{a}{Column 
(1) name,   
(2) redshift, 
(3) averaged Cousins R band surface brightness within $r_{e}$, extinction and K corrected, using Equation (2),
(4) Extinction correction from Schlegel et al. (1998),            
(5) K correction from Poggianti et al. (1997), 
(6) observed host galaxy magnitude in the Cousins R band from Urry et al. (2000),
(7) error of host galaxy magnitude,
(8) half light radius from Urry et al. (2000),    
(9) error of $r_{e}$,    
(10) stellar velocity dispersion,
(11) fitting error of velocity dispersion,
(12) observing run,
(13) exposure time,
(14) extraction radius,
(15) signal-to-noise ratio per pixel, measured at 6000 \AA~ in each combined galaxy spectrum,
(16) correction factor for velocity dispersions to a 3.4$^{\prime\prime}$ aperture at the distance of the Coma cluster.
Due to the presence of the bright AGN continuum, the observed S/N ratio is much higher than
the actual S/N of the galaxy absorption lines.
}
\tablerefs{
a) total galaxy magnitude, $r_{e}$, and $<\mu_{e}>$ of these radio galaxies are from Dunlop et al. (2003).
}
\end{deluxetable}

\begin{deluxetable}{lcrrrrrrrrrr}
\tablewidth{0pt}
\tablecaption{Previous measurements from the literature}
\tablehead{
\colhead{Name}  &
\colhead{z}     &
\colhead{$<\mu_{e}>$} &
\colhead{$A_R$} &
\colhead{$K_R$} &
\colhead{$m_R$} & 
\colhead{$\Delta m_R$} & 
\colhead{$r_e$} &
\colhead{$\Delta r_e$}    &
\colhead{$\sigma$}        &
\colhead{$\Delta \sigma$} &
\colhead{ref.}            \\
\colhead{} &
\colhead{} &
\colhead{} &
\colhead{} &
\colhead{} &
\colhead{} &
\colhead{} & 
\colhead{$^{\prime\prime}$} &
\colhead{$^{\prime\prime}$} &
\colhead{km sec$^{-1}$} &
\colhead{km sec$^{-1}$} &
\colhead{ } \\
\colhead{1} &
\colhead{2} &
\colhead{3} &
\colhead{4} &
\colhead{5} &
\colhead{6} & 
\colhead{7} &
\colhead{8} &
\colhead{9} &
\colhead{10} &
\colhead{11} &
\colhead{12}   
} 
\tablecolumns{12}
\startdata
Mrk501   &  0.034& 20.40&  0.05& 0.036 &13.07$^{a}$& -&     12.99$^{b}$ & 4.15&  291.&   13. &F  \\
Mrk421   &  0.031& 20.29&  0.04& 0.033 &13.29& 0.02&        11.04$^{b}$ & 3.49&  230.&    8. &BF \\
Mrk180   &  0.045& 20.40&  0.04& 0.048 &14.45& 0.02&        7.02$^{b}$  & 2.22&  230.&    8. &BF \\
3C371    &  0.051& 20.61&  0.10& 0.054 &13.87& 0.02&        10.53$^{b}$ & 1.09&  272.&   16. &BF \\
1Zw187   &  0.055& 19.61&  0.08& 0.059 &15.49& 0.02&                3.15& 0.05&  179.&   13. &B  \\
0521-365 &  0.055& 18.44&  0.10& 0.059 &14.60& 0.01&                2.80& 0.07&  264.&   16. &BF \\
0548-322 &  0.069& 20.40&  0.09& 0.076 &14.62& 0.01&                7.05& 0.15&  233.&   14. &BF \\
0706+591 &  0.125& 19.59&  0.10& 0.156 &15.94& 0.01&                3.05& 0.07&  233.&   25. &B  \\
1959+650 &  0.048& 19.73&  0.47& 0.051 &14.92& 0.02&                5.10& 0.10&  195.&   15. &F  \\
2201+044 &  0.027& 19.63&  0.11& 0.029 &13.74& 0.01&                6.78& 0.08&  178.&    6. &BF \\
2344+514 &  0.044& 19.06&  0.58& 0.046 &14.01& 0.01&                5.93& 0.02&  305.&   25. &B  \\
\enddata
\label{previ}          
\tablenotetext{a}{Column 
(1) name,   
(2) redshift, 
(3) average surface brightness within $r_{e}$, extinction and K corrected, using equation (2),
(4) extinction correction from Schlegel et al. (1998),            
(5) K correction from Poggianti et al. (1997), 
(6) host galaxy magnitude in the Cousins R band from Urry et al. (2000),
(7) error in host galaxy magnitude,
(8) half light radius in arcsec from Urry et al. (2000),    
(9) error of $r_{e}$ in arcsec,    
(10) velocity dispersion, corrected to an aperture of $3.4^{\prime\prime}$ at the Coma cluster distance,
(11) dispersion error in km sec$^{-1}$, 
(12) reference for dispersion measurements (B: Barth et al. (2002), F: Falomo et al. (2003), BF: log average of Barth et al. (2002)
and Falomo et al. (2003)).
}
\tablerefs{ 
a) total galaxy magnitude is from Barth et al. (2003),
b) $r_{e}$ is the median of published $r_{e}$ values from Falomo et al. 2003.}
\end{deluxetable}

\end{document}